\documentclass[preprint,showpacs,preprintnumbers,amsmath,amssymb]{revtex4}

\usepackage{epsfig}
\usepackage{graphicx}
\usepackage{dcolumn}
\usepackage{bm}

\newcommand{\ord}{{\cal O}}
\def\beq{\begin{equation}}
\def\eeq#1{\label{#1}\end{equation}}
\def\eeqn{\end{equation}}
\newcommand\iden{\leavevmode\hbox{\small1\normalsize\kern-.33em1}}
\newcommand{\be}{\begin{equation}}
\newcommand{\ee}{\end{equation}}
\newcommand{\bea}{\begin{eqnarray}}
\newcommand{\eea}{\end{eqnarray}}

\let\jnfont=\rm
\def\NPB#1,{{\jnfont Nucl.\ Phys.\ B }{\bf #1},}
\def\PLB#1,{{\jnfont Phys.\ Lett.\ B }{\bf #1},}
\def\EPJC#1,{{\jnfont Eur.\ Phys.\ Jour.\ C }{\bf #1},}
\def\PRD#1,{{\jnfont Phys.\ Rev.\ D }{\bf #1},}
\def\PRL#1,{{\jnfont Phys.\ Rev.\ Lett.\ }{\bf #1},}
\def\MPLA#1,{{\jnfont Mod.\ Phys.\ Lett.\ A }{\bf #1},}
\def\JPG#1,{{\jnfont J.\ Phys.\ G }{\bf #1},}
\def\CTP#1,{{\jnfont Commun.\ Theor.\ Phys.\ }{\bf #1},}
\def\JHEP#1,{{\jnfont JHEP \ }{\bf #1},}
\def\NPPS#1,{{\jnfont Nucl.\ Phys.\ Proc.\ Suppl.\ }{\bf #1},}
\def\CPC#1,{{\jnfont Computl.\ Phys.\ Commun.\ }{\bf #1},}
\def\FP#1,{{\jnfont Fortsch.\ Phys.\ }{\bf #1},}

\begin{document}

\preprint{\parbox{1.2in}{\noindent }}

\title{\ \\[10mm]  Higgs boson decays and production via gluon fusion at LHC
                   in littlest Higgs models with T-parity}

\author{\ \\[2mm] Lei Wang, Jin Min Yang \\ ~}

\affiliation{Institute of Theoretical Physics, Academia Sinica,
             Beijing 100190, China \vspace*{1.5cm}}

\begin{abstract}
We study the Higgs boson decays and production via gluon fusion
at the LHC as a probe of two typical littlest Higgs models which
introduce a top quark partner with different (even and odd) T-parity
to cancel the Higgs mass quadratic divergence contributed by
the top quark. For each model we consider two different choices for
the  down-type quark Yukawa couplings.
We first examine the branching ratios of the Higgs boson decays
and then study the production via gluon fusion followed by
the decay into two photons or two weak gauge bosons.
We find that the predictions can be quite different for
different models or different choices of down-type quark
Yukawa couplings and all these predictions can sizably deviate
from the SM predictions. So the Higgs boson processes at the LHC
can be a sensitive probe for these littlest Higgs models.
\vspace*{1cm}
\end{abstract}


\maketitle

\section{Introduction}
To solve the fine-tuning problem of the Standard Model (SM), the
little Higgs \cite{ref1} is proposed as a kind of
electroweak symmetry breaking mechanism accomplished by a naturally
light Higgs sector.  The Higgs boson remains light, being protected by
the approximate global symmetry and free from one-loop quadratic sensitivity
to the cutoff scale.
The littlest Higgs model \cite{ref2} provides an economical approach
which implements the idea of the little Higgs.
However, due to the tree-level mixing of heavy and light mass
eigenstates, the electroweak precision tests can give strong
constraints on this model \cite{ref3}, which would require raising
the mass scale of the new particles to be much higher than TeV and
thus reintroduce the fine-tuning in the Higgs potential \cite{ref4}.
To tackle this problem, a discrete symmetry called T-parity is
proposed \cite{ref5}, which forbids those tree-level contributions
to the electroweak observables.
In the pioneer version of this model (hereafter called model-I)
\cite{ref5}, the T-parity is simply implemented by adding the
T-parity images for the original top quark interaction to make the
Lagrangian T-invariant. A characteristic prediction of this model is
a T-even top partner which cancels the Higgs mass quadratic
divergence contributed by the top quark.

An alternative implementation of T-parity has recently been
proposed (hereafter called model-II)  \cite{ref6},  where all new
particles including the heavy top partner responsible for
cancelling the SM one-loop quadratic divergence are odd under
T-parity. An obvious virtue of this model is that the spectrum of
the third-generation quark sector is simplified \cite{ref6}.

These littlest Higgs models with T-parity (LHT) mainly alter the
property of the Higgs boson and hence the hints of these models
can be unravelled from various Higgs boson processes
\cite{ref12,yuanlone,invisible,beforewang,xfhan}. Since different
models have different predictions for Higgs boson processes, it is
important to study the models comparatively. In this work we
perform such a comparative study for model-I and model-II focusing
on the decay branching ratios of the Higgs boson as well as the
production at the LHC via gluon fusion  followed by the decay into
two photons or two weak gauge bosons. Since both models can have
two different choices for the down-type quark Yukawa couplings, we
will consider the two choices for each model. In our analysis we
will show the predictions of these two models and compare with the
SM results.

This work is organized as follows. In Sec. II we recapitulate the LHT
models with emphasis on model-II since model-I has been intensively
discussed in the literature.
Then we perform a comparative study for model-I and
model-II focusing on the decay branching ratios of the Higgs boson
in Sec. III and the production at the LHC via gluon fusion followed by
the decay into two photons or two weak gauge bosons in Sec. IV.
Finally, we give our conclusion in Sec. V.

\section{The littlest Higgs model with T-parity}

The original littlest Higgs model \cite{ref2} is based on a
non-linear sigma model describing the spontaneous breaking of a
global $SU(5)$ down to a global $SO(5)$ at an energy
scale $f\sim\ord({\rm TeV})$. The vacuum expectation value of an
$SU(5)$ symmetric tensor $\Sigma$ is proportional to \beq \Sigma_0
\,=\, \left(\begin{array}{ccc}
0& 0& \iden\\
0& 1& 0\\
\iden& 0& 0\\
\end{array}\right),
\eeq{sigma0} where $\iden$ represents a unit $2\times 2$ matrix. The
low energy dynamics of the non-linear sigma model is described in terms of the
field
\beq
\Sigma(x) \,=\, e^{i \Pi/f} \Sigma_0 e^{i \Pi^T/f} \,=\, e^{2i \Pi/f} \Sigma_0
\eeq{sigma_def}
where
\beq \Pi(x)=\sum_{a=1}^{14} \pi^a(x) X^a,
 \eeq{pionmatrix}
with $\pi^a(x)$ being the Goldstone fields corresponding to 14
broken generators $X^a$ for the $SU(5)\to SO(5)$ breaking.

In the pioneer version of littlest Higgs model with T-parity
(model-I), the T-parity in the top quark sector is implemented by
simply adding the T-parity images of the original interaction to
make the Lagrangian T-invariant. Thus, it predicts a T-even
top partner which cancels the Higgs mass quadratic divergence
contributed by the top quark. Since there are detailed descriptions
for this model in the literature \cite{ref5,ref12}, we do not discuss
it in detail here. In the following we recapitulate an alternative
version of T-parity construction (model-II) \cite{ref6}.

In model-II, to implement T-parity in the fermion sector, it
requires the introduction of the mirror fermions. For each SM
lepton/quark doublet, under the $SU(2)_1\times SU(2)_2$ gauge
symmetry, two fermion doublets $q_1(2,1)$ and $q_2 ({1,2})$ are
introduced. They can be embedded into the incomplete representation
multiplets $\Psi_1$ and $\Psi_2$ of $SU(5)$. A right-handed $SO(5)$
multiplets  $\Psi_R$ transforming nonlinearly under the full $SU(5)$
is introduced to give mass to the extra fermions.  The field content
can be expressed as
\begin{equation}
\begin{array}{ccc}
\Psi_1=\left(\begin{array}{c} q_1 \\ 0 \\ 0 \end{array}\right)\,,
& \Psi_2=\left(\begin{array}{c} 0 \\ 0 \\ q_2
\end{array}\right) \,,&
\Psi_R=\left(\begin{array}{c} \psi_R \\ \chi_R \\ \tilde{\psi}_R
\end{array}\right) ,
\end{array}
\end{equation}
where $q_A=-\sigma_2(u_{L_A},d_{L_A})^T=(id_{L_A},-iu_{L_A})^T$ with
$A=1,2$ and $\tilde{\psi}_R=\left( i d'_{R}, -i u'_{R} \right)^{\rm
T}$.  The second component of $\psi_R$ is $ -iq_R$. The mirror
fermions can obtain $\ord(f)$ mass via
\begin{equation} \label{eq5}
{\cal L}_{\kappa}=-\kappa_{ij}  f \left(\bar{\Psi}^i_2 \xi
+\bar{\Psi}^i_1 \Sigma_0\xi^\dagger \right)\Psi^j_R +h.c.,
\end{equation}
where $\xi=e^{i\Pi/f}$, $\Omega \equiv {\rm diag}(1,1,-1,1,1)$, and
$i,j=1,2,3$ are the generation indices. For simplicity, we assume
flavor diagonal and hence we have a universal $\kappa$ in our study.
The fields transform under $SU(5)$ as
\beq
  \Psi_1 \rightarrow V^* \Psi_1\,
  ~ \Psi_2 \rightarrow V \Psi_2\,, ~ \Psi_R \rightarrow U\Psi_R,
  ~ \xi \rightarrow V\xi U^\dagger=U\xi\Sigma_0 V^{\rm T}\Sigma_0,
  ~ \Sigma \rightarrow V\Sigma V^{\rm T} \, ,
\eeq{su5}
where $V$ denotes the $SU(5)$ rotation, and $U$ is the unbroken
$SO(5)$ rotation and is a
non-linear representation of the $SU(5)$. Under T-parity the
transformation laws are defined as
\beq
 \Psi_1 \rightarrow -\Omega\Sigma_0 \Psi_2\, ,
 ~\Psi_R \rightarrow -\Omega\Psi_R\, ,
 ~\xi \rightarrow \Omega \xi^\dagger \Omega\, ,
\eeq
where the transformation of $\Psi_1$  follows \cite{0811.2891}, and Eq.
(\ref{eq5}) has the full $SU(5)$ global symmetry and thus we have
$q_1\rightarrow -q_2$ and $\Sigma \rightarrow
\tilde{\Sigma}=\Sigma_0 \Omega \Sigma^\dagger \Omega \Sigma_0$ under
T-parity. Under the above transformations, the Lagrangian is
T-invariant.

The Lagrangian in Eq. (\ref{eq5}) contains the new Higgs boson interactions
and the mass terms. For the first and second generations we have
\begin{eqnarray} \label{eq8}
{\cal L}_{\kappa} &\simeq&-\sqrt{2} \kappa f \left[\bar{d}_{L_-}
d'_{R}+\frac{1+c_\xi}{2} \bar{u}_{L_-} u'_R
-\frac{1-c_\xi}{2}\bar{u}_{L_-}q_R
+\frac{s_\xi}{\sqrt{2}}\bar{u}_{L_+} \chi_R \right] +{\rm h.c.}\, ,
 \end{eqnarray}
where we ignored the generation indices, and
 $c_\xi \equiv \cos\frac{v+h}{\sqrt{2}f}$ and
$s_\xi \equiv \sin\frac{v+h}{\sqrt{2}f}$ come from the non-linear
sigma model field $\xi$, with  $h$ and $v$ being the neutral Higgs
boson field and its vacuum expectation value, respectively. The
fermion $u_{L_{-}} = (u_{L_1}+ u_{L_2})/\sqrt{2}$ is T-odd, which
together with $u'_R$ gets a mass, and $u_{L_{+}} = (u_{L_1}-
u_{L_2})/\sqrt{2}$ is T-even and massless. The same definition also
applies to the down-type quarks. The fields $q_R$ and $\chi_R$ can
obtain large Dirac masses by introducing additional fields, as
discussed in \cite{ref5}. In our study we assume both masses are 3.5
TeV. From Eq. (\ref{eq8}) we can see that the first component of the
doublet $\tilde{\psi}_R$ does not appear and the T-odd down-type
quarks have no tree-level coupling with the Higgs boson.

For the top quark interaction sector, in order to cancel the
quadratic divergence of the Higgs mass induced by the top quark, it
requires the introduction of additional singlets $U_1$ and $U_2$.
One can write down their interaction Lagrangian as \cite{ref6}
\begin{equation} \label{eq9}
{\cal L}_t= -\frac{\lambda}{2\sqrt{2}}f\epsilon_{ijk} \epsilon_{xy}
\left[(\bar{Q}_1)_i \Sigma_{jx} \Sigma_{ky}U_{R_1}-
(\bar{Q}_2\Sigma_0\Omega)_i \tilde{\Sigma}_{jx}
\tilde{\Sigma}_{ky}U_{R_2}\right] +{\rm h.c.} \, ,
\end{equation}
where the indices $i,j,k$ run from 1 to 3 whereas $x,y=4,5$, and
$Q_1=(q_1,U_{1},0_2)^{\rm T}$ whereas $Q_2=(0_2,U_{2},q_2)^{\rm T}$.
Under T-parity these fields transform as $Q_1 \rightarrow
-\Omega\Sigma_0 Q_2,~ U_{R_1}\rightarrow U_{R_2}$. Therefore, the
T-parity eigenstates are defined as $U_{L_{-}} =
(U_{1}-U_{2})/\sqrt{2}$ (T-odd) and $U_{L_{+}} = (U_{1}+
U_{2})/\sqrt{2}$ (T-even), and the same definition also applies to
the right-handed singlet. Eq. (\ref{eq9}) will introduce mixing
between the light T-even and the heavy T-even fermions, which can be
removed by the additional interactions:
\begin{equation} \label{eq10}
{\cal L}'_t= -\frac{\lambda'}{2\sqrt{2}}f\epsilon_{lmn}
\epsilon_{rs}\left[(\bar{Q}_2)_l \Sigma'_{mr}
\Sigma'_{ns}U_{R_1}-(\bar{Q}_1\Omega\Sigma_0)_l \tilde{\Sigma}'_{mr}
\tilde{\Sigma}'_{ns}U_{R_2}\right] +{\rm h.c.} \, ,
\end{equation}
where $\Sigma'=\Omega \Sigma^\dagger \Omega$,
$\Sigma'\rightarrow\tilde{\Sigma}'=\Sigma_0\Sigma\Sigma_0$ under T-parity,
and the indices $l$, $m$, $n$ run from 3 to 5 whereas $r$, $s$=1,2.
Adding ${\cal L'}_t$ to ${\cal L}_t$ and taking
$\lambda'=\lambda$, we can get the following simple expression for
the top quark Yukawa coupling sector
\begin{eqnarray}
{\cal L}_t-{\cal L}'_t&\simeq& -\lambda f \left(s_\Sigma
\bar{u}_{L_+}U_{R_+}+\frac{1+c_\Sigma}{\sqrt{2}} \bar{U}_{L_-}
U_{R_-} \right)+{\rm h.c.}\, , \label{eq11}
\end{eqnarray}
where $c_\Sigma \equiv\cos\frac{\sqrt{2}(v+h)}{f}$ and
$s_\Sigma\equiv \sin\frac{\sqrt{2}(v+h)}{f}$ arise from the
non-linear sigma model field $\Sigma$. The field $U_{L_+}$  together
with $\chi_R$ gets a Dirac mass in Eq. (\ref{eq5}). From
Eq.(\ref{eq5}) we can get the Higgs boson interactions and the mass
terms for the third generation fermions
\begin{eqnarray}
{\cal L}_{\kappa} &\simeq&-\sqrt{2} \kappa f \left [ \bar{d}_{L_-}
d'_{R}+\frac{1+c_\xi}{2} \bar{u}_{L_-} u'_R
-\frac{1-c_\xi}{2}\bar{u}_{L_-}q_R \right. \nonumber \\
&&\left. -\frac{s_\xi}{\sqrt{2}}\bar{U}_{L_-}q_R
  -\frac{s_\xi}{\sqrt{2}}\bar{U}_{L_-}u'_R
  +\frac{s_\xi}{\sqrt{2}}\bar{u}_{L_+} \chi_R
  +c_\xi\bar{U}_{L_+}\chi_R \right] +{\rm h.c.}. \label{eq12}
 \end{eqnarray}
The Yukawa couplings of up-type quarks for the first and second
generations are given by a similar Lagrangian as for the top quark,
but without introducing any extra singlet fields:
\begin{equation} \label{eq13}
{\cal L}_u= -\frac{\lambda_u}{2\sqrt{2}}f\epsilon_{ijk}
\epsilon_{xy} \left[(\bar{\Psi}_1)_i \Sigma_{jx} \Sigma_{ky}-
(\bar{\Psi}_2\Sigma_0\Omega)_i \tilde{\Sigma}_{jx}
\tilde{\Sigma}_{ky}\right]u_{R}+{\rm h.c.}\, ,
\end{equation}
where $u_R\rightarrow u_R$ under T-parity. Eq.(\ref{eq13}) contains the
following Higgs boson interactions as well as the mass term for
up-type quarks of the first and second generations
\begin{equation} \label{eq14}
{\cal L}_u \simeq -\frac{\lambda_u}{\sqrt{2}} f s_\Sigma u_{L_+}u_R+{\rm h.c.} .
\end{equation}
After diagonalizing the mass matrix in
Eqs.(\ref{eq8},\ref{eq11},\ref{eq12},\ref{eq14}), we can get the
mass eigenstates and the Higgs couplings. For each SM fermion
doublet, there are $d_-$, $u_-$ , $q$ (T-odd) and $\chi$ (T-even).
Besides, the top quark has a T-odd  partner $T_-$ which cancels the
one loop quadratic divergence of Higgs mass induced by the top
quark.

Higgs boson has the couplings with other particles including
down-type quarks, leptons, SM gauge bosons, extra gauge bosons and
scalar particles. These couplings are same as in model-I and can be
found in \cite{ref12,taoh03}. Here we list the Higgs boson couplings
with the down-type quarks and the SM gauge bosons, normalized with
the corresponding couplings in the SM,
\begin{eqnarray}
\frac{g_{hd\bar{d}}}{g_{hd\bar{d}}^{\rm SM}}
&\approx&1-\frac{1}{4}\frac{v_{SM}^2}{f^2}+\frac{7}{32}
\frac{v_{SM}^4}{f^4} ~~~~{\rm for~Case~A}, \label{Higgs-downA} \nonumber\\
&\approx&1-\frac{5}{4}\frac{v_{SM}^2}{f^2}-\frac{17}{32}
  \frac{v_{SM}^4}{f^4} ~~~~{\rm for~Case~B},\nonumber\\
\frac{g_{hVV}}{g_{hVV}^{\rm SM}} &\approx&
  1-\frac{1}{4}\frac{v^2_{SM}}{f^2}-\frac{1}{32}\frac{v^4_{SM}}{f^4}, ~~(V=Z,W),
\label{eq15}
\end{eqnarray}
where
$G_{F}=1/(\sqrt{2}v_{sm}^2)$ with $v_{sm}=f\sqrt{1-\cos(\sqrt{2}v/f)}$.
The relation of down-type quark couplings also
applies to the lepton couplings.

\section{Higgs decay branching ratios in LHT models}
In both model-I and model-II, the heavy photon $A_H$ is the lightest
T-odd particle with a mass given by
\begin{equation}
M^2_{A_H} = \frac{{g^\prime}^2 f^2}{5}-\frac{{g^\prime}^2
v_{SM}^2}{4}.
\end{equation}
As discussed in \cite{ref18}, the scale $f$ in model-I may be as low
as 500 GeV, and the constraint in model-II is expected to be even
weaker \cite{ref6}.  For $f=500$ GeV, $A_H$ has a mass of about 65
GeV. Therefore, in addition to the SM decay channels, the new decay
$h\rightarrow A_H A_H$ will open for $m_h\geq 2m_{A_H}$ and the
partial width is given by \bea \Gamma(h \to A_{H} A_{H}) & = &
\frac{g_{hA_H A_H}^{2} m_h^3}{128 \pi m_{A_H}^4}
  \sqrt{1-\beta_{A_H}}\left(1-\beta_{A_H}+\frac{3}{4}\beta_{A_H}^2\right),
\eea where $\beta_{A_H}=4m_{A_H}^2/m_{h}^2$. Because $A_H$ is
stable, there are no off-shell decays $h\rightarrow A^*_{H}A^*_{H}$
or $h\rightarrow A_{H}A^*_{H}$. In the LHT models the partial widths
of the Higgs decays to the SM particles can be obtained as $\Gamma(h
\to XX)= \Gamma(h \to XX)_{SM}(g_{hXX}/g_{hXX}^{SM})^2$ ($X$ denotes
a SM particle), where $g_{hXX}/g_{hXX}^{SM}$ is predicted by the LHT
models and $\Gamma(h \to XX)_{SM}$ is calculated with the code
Hdecay \cite{hdecay} (the relevant higher order QCD and electroweak
corrections are considered in this code).

Note that in the LHT models the corrections to the tree-level decays
$h\to f \bar{f},WW,ZZ$ are mainly from the suppression of the
corresponding couplings. For the loop-induced decay $h \to gg$, in
addition to the top quark loops, the loops of new T-even and T-odd
quarks also come into play. For the decay $h \to Z\gamma$, the $W$
boson loop contribution is dominant \cite{invi19} and thus we only
consider the alteration of the Higgs coupling with the $W$ boson.
The decay channel $h \to \gamma\gamma$ is a focus of our discussion.
In addition to the contributions of top quark and $W$ boson, the new
charged heavy fermions, gauge bosons and scalar particles will
contribute to the decay $h \to \gamma\gamma$. Following the approach
in \cite{taohrr}, we calculate the partial decay width of  $h \to
\gamma\gamma$ at one-loop. Because the QCD radiative corrections are
rather small \cite{hdecay}, our result is precise enough.

In our calculation we take $\kappa=1$
and $\lambda'=\lambda$ in model-II, and $r=1$ in model-I.
Our calculations show that the results are not so sensitive to
these parameters, but very sensitive to the Higgs mass
$m_h$ and breaking scale $f$. We take $100 ~{\rm GeV}\leq m_h\leq 500~{\rm GeV}$
and $500~{\rm GeV} \leq f\leq 2 ~{\rm TeV}$.

\begin{figure}[htb]
\epsfig{file=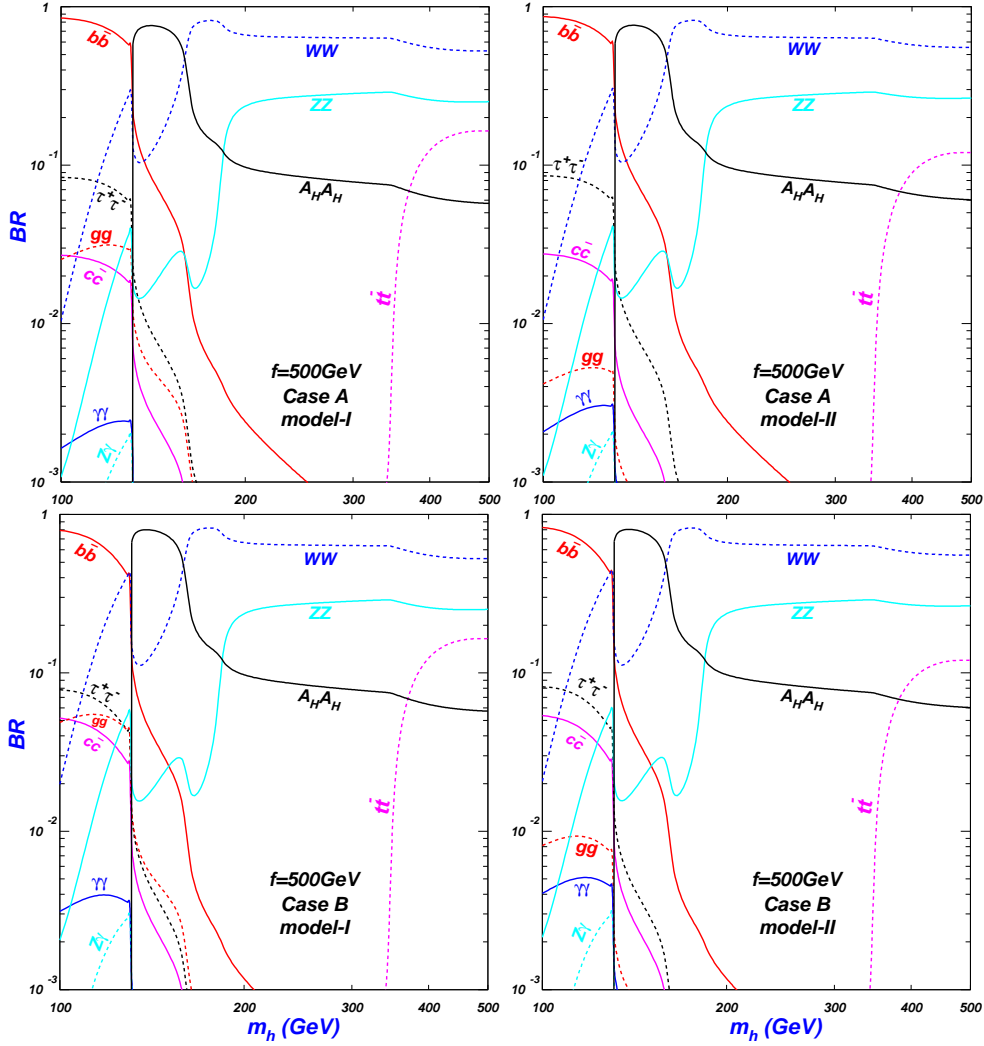,width=13cm}
\vspace*{-0.5cm}
\caption{\small The Higgs decay branching ratios versus the Higgs mass
           in model-I and model-II.}
\label{mh500}
 \end{figure}
\begin{figure}[htb]
\epsfig{file=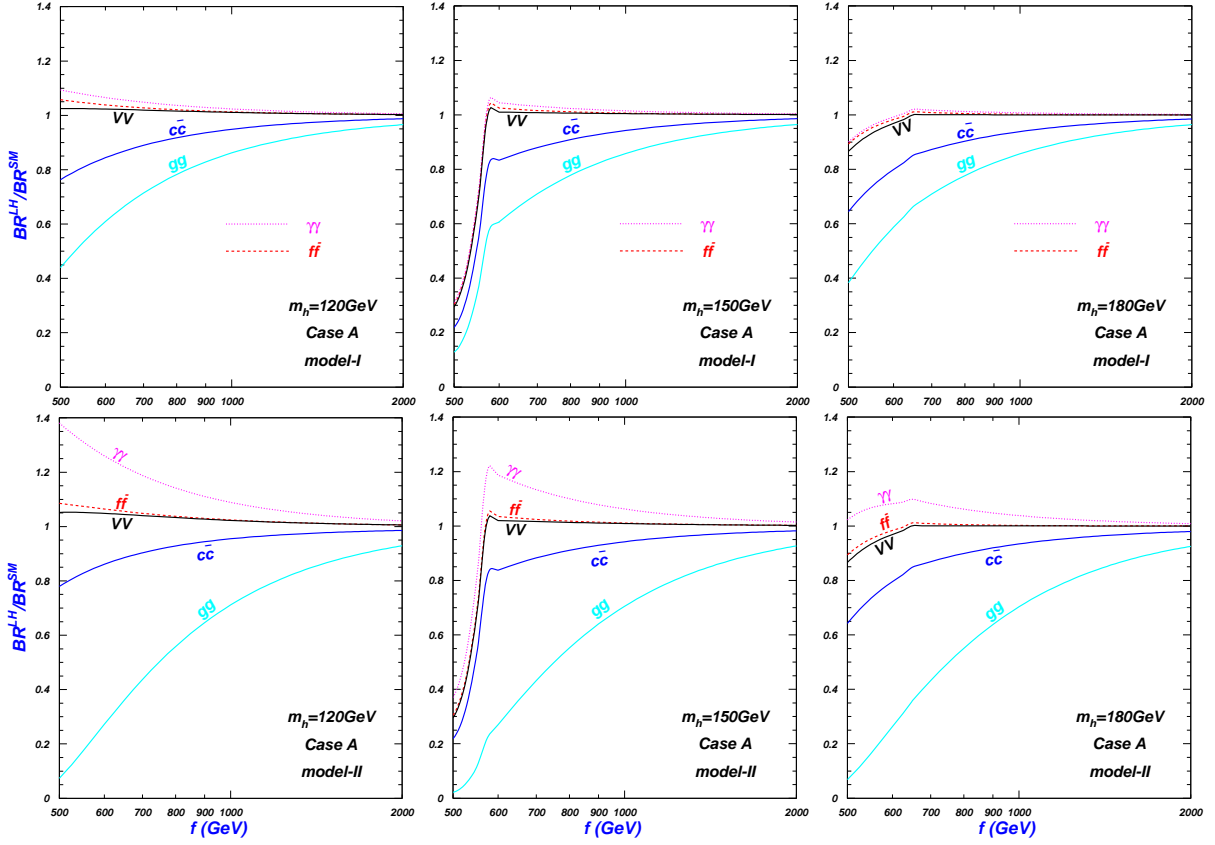,width=16cm}
\vspace*{-0.5cm}
\caption{\small  The Higgs decay branching ratios
normalized to the SM predictions in model-I and model-II.
For the decay channel $VV$, $V$ can be $Z$ or $W$, while for $f\bar f$,
$f$ denotes a down-type quark or lepton.}
\label{fa}
 \end{figure}
\begin{figure}[htb]
\epsfig{file=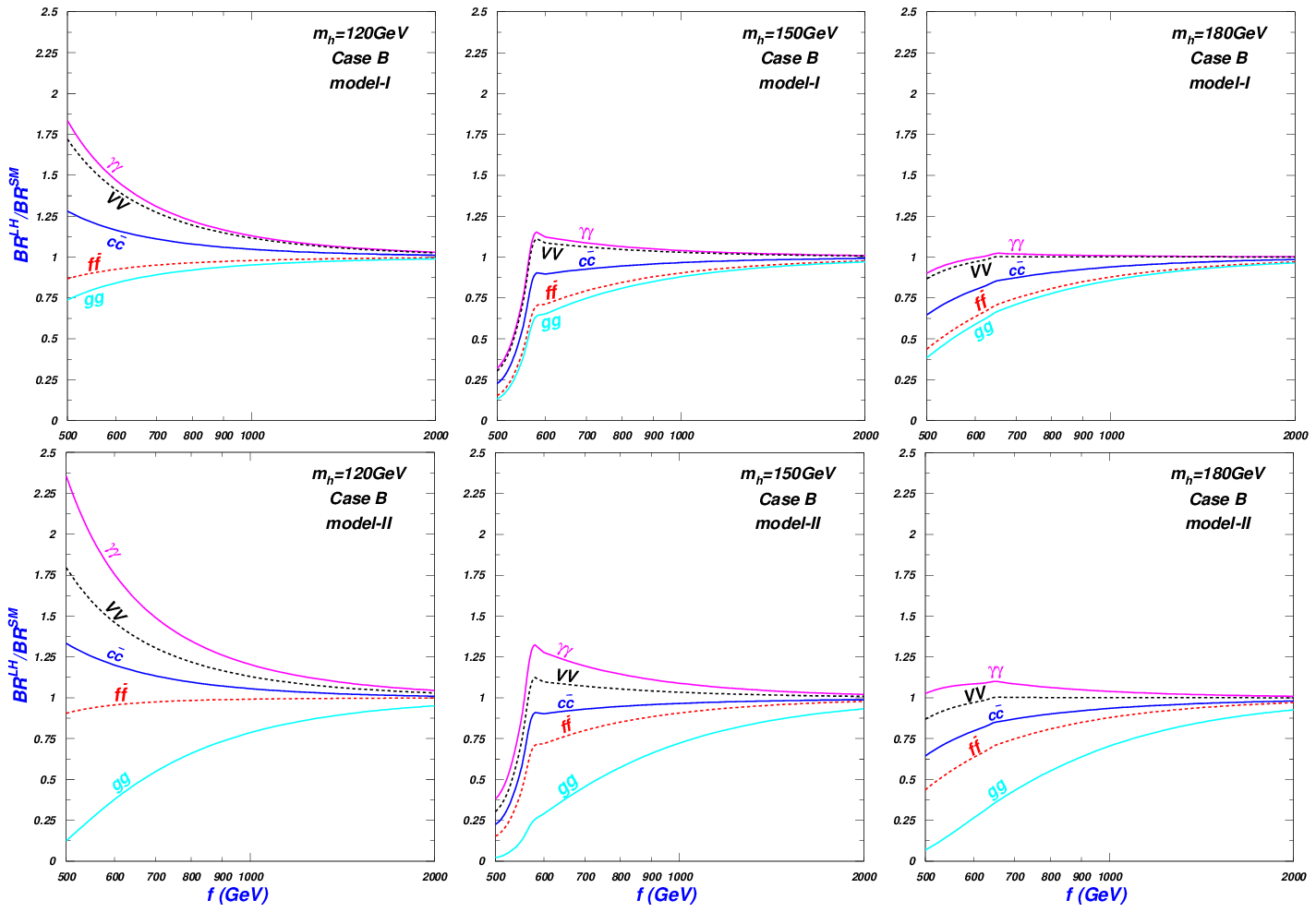,width=16cm}
\vspace*{-0.5cm}
\caption{\small Same as Fig.\ref{fa}, but for Case B.}
\label{fb}
 \end{figure}

In Fig.~\ref{mh500} we plot the Higgs decay branching ratios versus
the Higgs boson mass in model-I and model-II. Comparing the left
panels with the right panels, we see that the two models predict
very different branching ratios for $h\to gg$. Further, comparing
the upper panels with the lower panels, we see that for each model
the two cases give quite different branching ratios for $h\to gg$ or
$h\to \gamma\gamma$. In both models with $f=500$ GeV, the new decay
$h\to A_HA_H$ opens up for $m_h \geq 130$ GeV. This decay mode can
be dominant and over $70\%$ for $130 ~{\rm GeV}<m_h<150 ~{\rm GeV}$,
then it decreases as $m_h$ gets large and become comparable with $h
\to WW^*$ at $m_h \simeq 160$ GeV. The reason is that the Higgs
coupling with $A_H$ is of the electroweak strength and much
larger than the Yukawa coupling of $b$ quark. For $130 ~{\rm
GeV}<m_h<150 ~{\rm GeV}$, the decay width of $h\to A_HA_H$ is much
larger than the decay $h \to bb$ and the off-shell decay $h\to
W W^*$.  Here we fixed $f=500$ GeV and  did not show the
dependence on $f$. As $f$ gets larger, the decay $h\to A_HA_H$
becomes less important.

In Figs.~\ref{fa} and \ref{fb}, we plot the Higgs decay branching
ratios normalized to the SM predictions in model-I and model-II for
Case A and Case B, respectively. We see that for a small value of
$f$ the two models can predict quite different branching ratios from
the SM predictions. As $f$ gets large, the deviation from the SM
prediction for each decay mode becomes small and finally reduce to
the SM results when $f$ is up to 2 TeV. The deviation from the SM
prediction is also sensitive to the Higgs boson mass. Again, the
results show that the two models predict quite different branching
ratios for $h\to gg$ or $h\to \gamma\gamma$; while for other decay
modes the two models give the similar results. Besides, the
predictions of various branching ratios in Case A and Case B can be
sizably different for $m_h=120GeV$ and a small value of $f$.
In the following we give some explanations for the above features:
\begin{itemize}
\item[(1)] First we explain why the branching ratio of $h\to gg$
($h\to \gamma\gamma$) in model-II is smaller (larger) than in
model-I. The main contributions of model-I and model-II to the decay
$h\to gg$ are from the loops of the fermions whose couplings to $h$
are given by \small
\begin{eqnarray}
-\frac{m_t}{v}y_t\bar{t}th-\frac{m_T}{v}y_T\bar{T}Th+
\sum_{i=1}^{3}
-\frac{m_{u_{-}^i}}{v}y_{u_{-}^i}\bar{u}_{-}^iu_{-}^ih
-\frac{m_{q^i}}{v}y_{q^i}\bar{q}^iq^i h
-\frac{m_{\chi^i}}{v}y_{\chi^i}\bar{\chi}^i\chi^i h
\end{eqnarray}
\normalsize
in model-I and
\small
\begin{eqnarray}
-\frac{m_t}{v}y'_t\bar{t}th
-\frac{m_{_{U_{-}}}}{v}y'_{_{U_{-}}}\bar{U}_{-}U_{-}h+
\sum_{i=1}^{3}
-\frac{m_{u_{-}^i}}{v}y'_{u_{-}^i}\bar{u}_{-}^iu_{-}^i
h-\frac{m_{q^i}}{v}y'_{q^i}\bar{q}^iq^i h
-\frac{m_{\chi^i}}{v}y'_{\chi^i}\bar{\chi}^i\chi^i h
\end{eqnarray}
\normalsize in model-II. Here all the particles are the mass
eigenstates (the diagonalization of the mass matrix was performed
numerically in our analysis). The contributions of these fermion
loops are not sensitive to the mass values when the fermion masses
are much larger than half of the Higgs boson mass. Hence, the
contributions of model-I and model-II  are approximately
proportional to $y^2_{_{I}}/v^2$ and $y^2_{_{II}}/v^2$, where
$y_{_{I}}$ and $y_{_{II}}$ denote the sum of $y$ and $y'$ in Eqs.
(18) and (19), respectively. As $y^2_{_{II}}$ is smaller than
$y^2_{_{I}}$ in the parameter space we chose (for example, Table 1
shows the values of $y_{_{I}}$ and $y_{_{II}}$ for $f=700$ GeV),
the decay width of $h\to gg$ in model-II is thus smaller than in
model-I. For the decay $h\to \gamma\gamma$, besides the fermion
loops, the boson loops also contribute but with an opposite sign.
While the fermion loop contribution in model-II is smaller than in
model-I, the contributions of boson loops are equal in both
models. The extent of cancellation between fermion and boson loops
in model-I is more severe than in model-II. Thus, the decay width
of $h \to \gamma\gamma$ in mode-II is larger than in model-I. On
the other hand, for the total decay width of the Higgs boson,
depending on the value of the Higgs boson mass, it can be
dominated by the decay $h\to b\bar{b}$, $h \to VV$ or $h \to A_H
A_H$, and each of these decays has the same width in both models.
Therefore, the branching ratio of $h\to gg$ ($h\to \gamma\gamma$)
in model-II is smaller (larger) than that in model-I.
\begin{table}[t]
\caption{The values of $y$ and $y'$ in Eqs. (18) and (19) for $f=700$ GeV.}
\begin{center}
\begin{tabular}{|c|c|c|c|c|c|}
\hline $y_t$ & $y_T$ & $y_{u_{-}^1}+y_{u_{-}^2}+y_{u_{-}^3}$&
$y_{q_{-}^1}+y_{q_{-}^2}+y_{q_{-}^3}$&
 $y_{\chi_{-}^1}+y_{\chi_{-}^2}+y_{\chi_{-}^3}$& $y_{_{I}}$\\
 \hline
$0.947$ & $-0.036$& $-0.104$&
 $0.008$& $0$&0.815\\
\hline \hline $y'_t$ & $y'_{_{U_{-}}}$ &
$y'_{u_{-}^1}+y'_{u_{-}^2}+y'_{u_{-}^3}$&
$y'_{q_{-}^1}+y'_{q_{-}^2}+y'_{q_{-}^3}$&
 $y'_{\chi_{-}^1}+y'_{\chi_{-}^2}+y'_{\chi_{-}^3}$& $y_{_{II}}$\\
 \hline
$0.876$ & $-0.169$& $-0.057$&
 $0.001$& $-0.026$&0.625\\
\hline
\end{tabular}
\end{center}
\label{discovery_pot}
\end{table}
\item[(2)] The reason why the two models give similar results for
the decay modes of $c\bar c$ and $f\bar f$ ($f$ is a down-type quark
or lepton) is that the two models give respectively the same prediction
for the Higgs Yukawa couplings with down-type quarks, leptons, and almost
the same prediction for up-type quarks except top quark. Therefore, the
branching ratios of these decay channels are similar in both models.
\item[(3)] The $hb\bar{b}$ coupling in Case B is
           more suppressed than in Case A, as shown in Eq. (15).
For $m_h=120$ GeV, the decay $h \to b\bar{b}$ is dominant
and the total decay width in Case B is much smaller than
in Case A. Therefore, the predictions of various branching
ratios in Case B can be sizably different from those in Case A
for $m_h=120$ GeV and a small value of $f$.
\end{itemize}

\section{The rate $\sigma(gg\to h) \times BR(h\to \gamma\gamma ~{\rm or~} VV)$
         at LHC in LHT models}

In the SM the Higgs productions at the LHC are dominated by gluon
fusion process. The $h \to \gamma\gamma$ channel shows very good
sensitivity in the range $114 {\rm ~GeV}<m_h<140 {\rm ~GeV}$.
For $2m_W<m_h<2m_Z$,
the decay $h \to WW \to l\nu l\nu$ provides the most sensitive search
channel. For $m_H > 130$ GeV (except the
interval between $2m_W$ and $2m_Z$), the channel $h \to ZZ^* \to 4l$
provides excellent sensitivity \cite{hsearch}. Especially, for
the  channel $h \to \gamma\gamma$, with an integrated luminosity 100 fb$^{-1}$
(10 fb$^{-1}$) from both ATLAS and CMS,
the rate $\sigma(gg \to h)\times BR(h \to \gamma\gamma)$ can be
measured to $10\%$ ($30\%$) \cite{yuanlone,lone9}. Once we find a
light Higgs boson at the LHC, this channel can provide a test
for different models. In the LHT models,
$\sigma(gg \to h)$ is strongly correlated with
$\Gamma(h \to gg)$, which depend on the same effective coupling
$hgg$.  In our results we use $\sigma(gg \to h)$ to denote
the hadronic cross section of the Higgs production proceeding
through  $gg \to h$ at parton level.
We use CTEQ6L \cite{cteq} for parton distributions,
with the renormalization scale $\mu_R$ and factorization
scale $\mu_F$ chosen to be $\mu_R=\mu_F=m_h$.

\begin{figure}[htb]
\epsfig{file=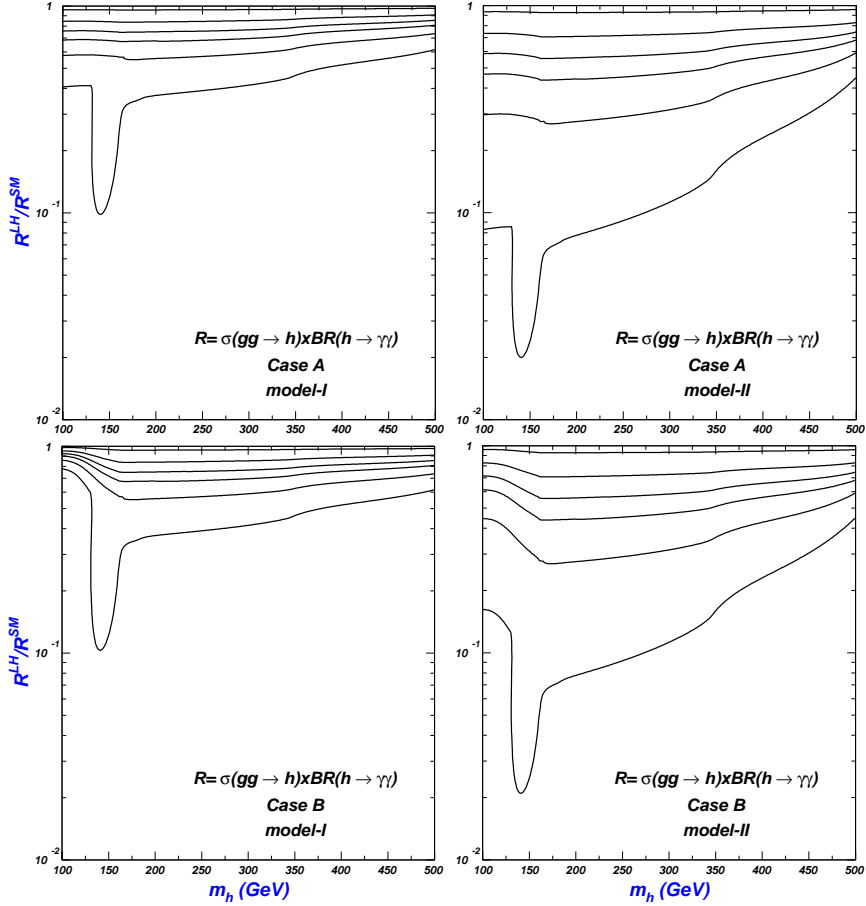,width=11.5cm}
\vspace*{-0.7cm}
\caption{The value of $\sigma(gg \to h) \times BR(h\to \gamma\gamma)$
normalized to the SM prediction in model-I and model-II.
The curves from bottom to top correspond to $f=500$ GeV,  600 GeV,
700 GeV, 800 GeV, 1 TeV, 2 TeV, respectively.
The cross section $\sigma(gg \to h)$ denotes the hadronic cross section
proceeding through $gg \to h$.}
\label{mhr}
\end{figure}
\begin{figure}[htb]
\vspace{0.2cm} \epsfig{file=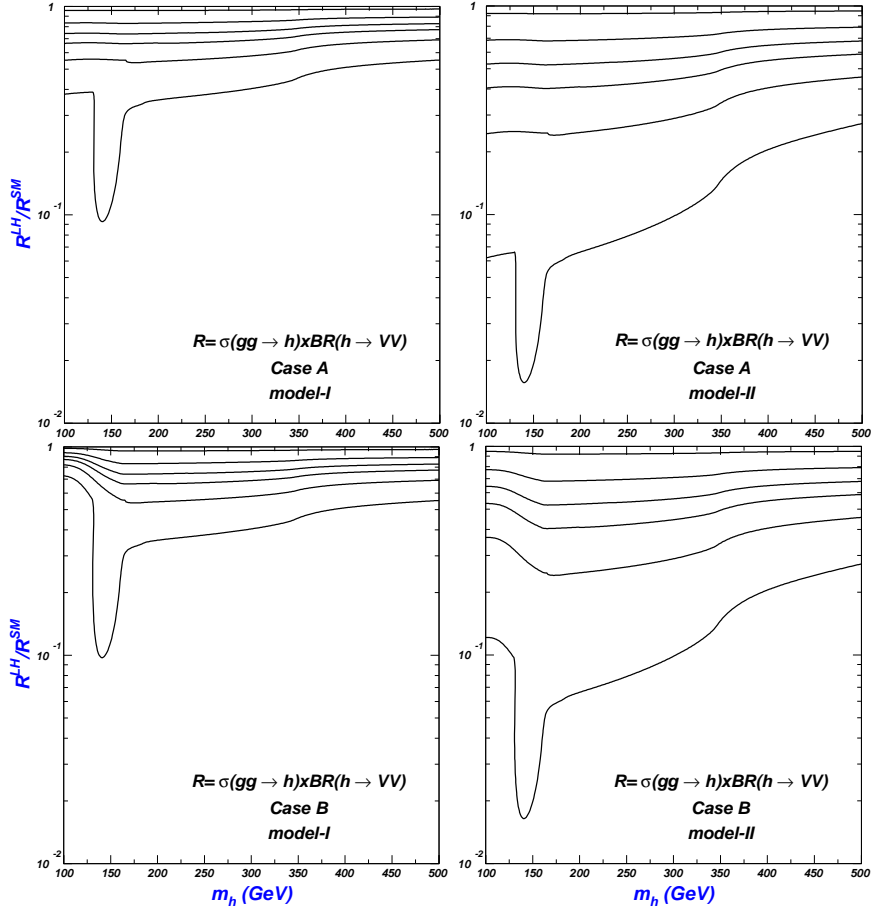,width=11.5cm}
\vspace*{-0.7cm}
\caption{Same as Fig. \ref{mhr}, but for
$\sigma(gg \to h) \times Br(h\to VV)$ ($V=Z, W$).}
\label{mhv}
\end{figure}

In Figs.~\ref{mhr} and \ref{mhv} we plot respectively the rates of
$\sigma(gg \to h) \times BR(h\to \gamma\gamma)$ and $\sigma(gg \to
h) \times BR(h\to VV)$ ($V=Z, W$) normalized to the SM predictions
in model-I and model-II. We see that compared with the SM
predictions the LHT models can suppress the rates sizably for a
small value of $f$. As $f$ gets large, the suppression is weakened
and finally the results reduce to the SM predictions for a
sufficiently large $f$ (about 2 TeV). Further, the two models can
give very different predictions. For example,  for $m_h=150$ GeV
and $f=500$ GeV, the rates are suppressed to $10^{-1}$ ($10^{-2}$)
relative to the SM predictions in model-I (model-II). The
deviation of the predictions between the two models can be
understood as follows. The production cross section  of $gg \to h$
is much smaller in model-II than in model-I since the process $gg
\to h$ is strongly correlated with the decay $h \to gg$ (the decay
width of $h \to gg$ in  model-II  is much smaller than in model-I,
as shown and explained in the preceding section). Although the
decay branching ratio of $h\to \gamma\gamma$ is larger in
model-II, the suppression of the production cross section of $gg
\to h$ is dominant and thus the rate $\sigma(gg \to h) \times
BR(h\to \gamma\gamma)$ is smaller in model-II. Besides,
Figs.~\ref{mhr} and \ref{mhv} showed that the predictions in Case
A and Case B can be sizably different in the range of $100
{\rm~GeV}\leq m_h\leq 130 {\rm ~GeV}$ for a small value of $f$.
The reason is that the two cases give quite different branching
ratios for $h\to \gamma\gamma$ or $h\to gg$ (and thus give
different cross sections for $gg\to h$), as shown and explained in
the preceding section.

\section{Conclusion}
In two typical littlest Higgs models which introduce a top
quark partner with different (even and odd) T-parity to cancel the
Higgs mass quadratic divergence contributed by the top quark, we
calculated the branching ratios of the Higgs boson decays and
examined the production at the LHC via gluon fusion followed by the
decay into two photons or two weak gauge bosons. For each model we
considered two different choices for the down-type quark Yukawa
couplings. From our numerical results we obtained the following
observations: (i) For the Higgs decays, we found that with $130 {\rm
~GeV}<m_h<150 {\rm ~GeV}$ and $f\simeq 500$ GeV, the new decay $h
\to A_H A_H$ can be the dominant mode. The two models can give very
different branching ratios from the SM predictions. Further, the
predictions between the two models can be quite different for
$BR(h\to \gamma\gamma)$ and $BR(h\to gg)$; while for other decay
modes both models give the similar predictions; (ii) For the rates
$\sigma(gg \to h) \times BR(h\to \gamma\gamma)$ and $\sigma(gg \to
h) \times BR(h\to VV)$ ($V=Z, W$) at the LHC, both models can give
severe suppression relative to the SM predictions, and the
predictions of the two models can also differ significantly; (iii)
For each model the two different choices for the  down-type quark
Yukawa couplings can also lead to different results. Therefore,
these Higgs processes at the LHC may be a sensitive probe for the
little Higgs theory and may even provide a way to distinguish the
different models or different scenarios.

\section*{Acknowledgement}
We thank C.-P. Yuan for discussions.
This work was supported  by the National Natural
Science Foundation of China (NNSFC) under Nos. 10821504,
10725526 and 10635030.

\end{document}